\documentclass[twocolumn,notitlepage,aps,pra]{revtex4-1}

\usepackage{algorithm}
\usepackage{algpseudocode}
\algrenewcommand{\algorithmiccomment}[1]{\# #1}

\newcommand{\ket}[1]{\left\vert#1\right\rangle}
\newcommand{\bra}[1]{\left\langle#1\right\vert}
\newcommand{\abs}[1]{\left\vert#1\right\vert}
\newcommand{\mb}[1]{{\bm#1}}
\usepackage{csvsimple}
\usepackage{datatool}
\usepackage{graphicx}
\usepackage{multirow}
\usepackage{amsmath}
\usepackage{amssymb}
\usepackage{amsthm}
\usepackage{eucal}
\usepackage{qcircuit}
\usepackage{bm}
\usepackage{upgreek}
\usepackage{mathrsfs}
\usepackage{color}
\usepackage{latexsym}
\usepackage[dvipsnames]{xcolor}
\usepackage{float}
\usepackage{enumerate}
\usepackage{stackengine}
\usepackage{braket}
\usepackage[titletoc]{appendix}
\usepackage{natbib}

\usepackage{xpatch}
\xpatchbibdriver{misc}
  {\printtext[parens]{\usebibmacro{date}}}
  {\iffieldundef{year}
    {}
    {\printtext[parens]{\usebibmacro{date}}}}
  {}
  {\typeout{There was an error patching biblatex-ieee (specifically, ieee.bbx's @online driver)}}

\theoremstyle{remark}

\usepackage{hyperref}
\hypersetup{colorlinks,linkcolor={blue},citecolor={red},urlcolor={blue}}

\newcommand{\bb}{\begin{equation}}
\newcommand{\ee}{\end{equation}}
\newcommand{\bbb}{\begin{equation*}}
\newcommand{\eee}{\end{equation*}}

\stackMath

\begin{document}

\title{Production of  photonic universal quantum gates enhanced by machine learning}

\author{Krishna Kumar Sabapathy}
\affiliation{Xanadu, 777 Bay Street, Toronto ON,  M5G 2C8, Canada}

\author{Haoyu Qi}
\affiliation{Xanadu, 777 Bay Street, Toronto ON,  M5G 2C8, Canada}

\author{Josh Izaac}
\affiliation{Xanadu, 777 Bay Street, Toronto ON,  M5G 2C8, Canada}

\author{Christian Weedbrook}
\affiliation{Xanadu, 777 Bay Street, Toronto ON,  M5G 2C8, Canada}

\begin{abstract}
We introduce photonic architectures for universal quantum computation. The first step is to produce a resource state which is a superposition of the first four Fock states with a probability $\geq 10^{-2}$, an increase by a factor of $10^4$ over standard sequential photon-subtraction techniques. The resource state is produced with near-perfect fidelity from a quantum gadget that uses displaced squeezed vacuum states, interferometers and photon-number resolving detectors. The parameters of this  gadget are trained using machine learning algorithms for variational circuits. We discuss in detail various aspects of the non-Gaussian state preparation resulting from the numerical experiments. We then propose a notion of resource farms where these gadgets are stacked in parallel, to increase the success probability further. We find a trade-off between the success probability of the farm, the error tolerance, and the number of gadgets. Using the resource states in conventional gate teleportation techniques we can then implement weak tuneable cubic phase gates.  The numerical tools that have been developed could potentially be useful for other applications in photonics as well. 
\end{abstract}

\maketitle 


\section{introduction}
Continuous-variable systems are one of the leading candidates for realizing universal quantum computation. In particular, there has been considerable recent progress in theory and experiments such as computation using temporally encoded modes~\cite{menicucci2006universal,yokoyama2015,larsen2018fibre} and frequency encoded domains~\cite{menicucci2008one}, loop-based architectures~\cite{motes2014scalable,qi2018linear,he2017time}, and bosonic-codes-based qubit/qumode encoding~\cite{chuang1997bosonic,bosonicterhal, PhysRevX.6.031006, PhysRevA.97.032323,PhysRevLett.119.030502,heeres2017implementing}, and promising applications in quantum key distribution~\cite{garcia2006unconditional,jouguet2013experimental,pirandola2015high} and sensing~\cite{giovannetti2011advances,pirandola2018advances}.

Universal quantum computation in continuous-variable systems requires non-Gaussian gates generated by Hamiltonians that are beyond quadratic in the quadrature operators~\cite{Lloyd1999a,Bartlett2002}.  In this context the quadrature phase gates which are of the form $\mathsf{\Theta}(\gamma) = \exp[i\gamma \hat{x}^n/\hbar]$ have played a very important role, especially the cubic phase gate corresponding to $n=3$, which has the lowest order non-Gaussian gate Hamiltonian among phase gates required for universal computation. Current methods to implement this gate are via gate teleportation where a resource state is prepared and used in a teleportation (measurement-based) circuit~\cite{Gottesman2001a,Sabapathy2018,Arzani2017,Marshall2015,Marek2018,Miyata2016a,Marek2011,Yukawa2013a}. However, the methods of preparation of the resource state either require very high photon-counting~\cite{Gottesman2001a,Ghose2007} or have a very low probability of success due to repeated photon subtractions, though some methods to improve the latter have been recently proposed~\cite{Honer2011,Marek2018a,Rosenblum2016}. 

We address this aspect of the resource state preparation wherein we constrain the resources to the simplest possible ones, namely, preparation of Gaussian multimode pure states followed by conditional photon detection measurements. For our purpose it suffices only to consider two-mode and three-mode architectures. We delegate the tuning of this multi-parameter constrained variational circuit to a machine learning algorithm that trains the circuit to learn the required state preparation, inspired by recent works~\cite{krenn2016automated,lau2016,melnikov2018active,gao2018experimental,cincio2018learning,arrazola2018machine}.  It turns out that we can prepare a learned state with near-perfect fidelity to the target state and with comparatively high probability, leading to a preparation efficiency greater than $1\%$, whereas sequential photon-subtraction techniques have an efficiency of $\sim 10^{-4}\%$ (see Appendix \ref{phosub}). We then estimate the resources required to build quantum resource farms  as a possible route to near-deterministic state generation, depending on an error threshold, for implementing  cubic phase gates. 

The rest of the paper is structured as follows. In Sec. \ref{gkp} we introduce the basic theory behind the gate teleportation method for implementing a weak-cubic phase gate using the requisite resource state. In Sec. \ref{ml} we introduce architectures for quantum gadgets whose parameters are trained to generate the resource states using machine learning algorithms. We also provide in detail the physical interpretation of the numerical results. We then obtain in Sec. \ref{qrf} a trade-off between the number of these quantum gadgets in what we term a quantum resource farm and the total success probability of producing the state. We then analyse the effects of photon loss in the quantum gadgets on the output states in Sec. \ref{noise}. We  conclude in Sec. \ref{con}. Explicit details of the numerical techniques is delegated to Appendix \ref{supp:num-tech}.

\section{Gottesman-Kitaev-Preskill (GKP) gate teleporation \label{gkp}}
We now focus on the optical implementation of the lowest-order quadrature phase gate, namely, the cubic phase gate that we denote by $\textsf{V}(\gamma) = \exp [i \gamma \hat{x}^3/\hbar]$, where $\gamma$ is the gate strength. All known methods to implement the cubic phase gate involve preparing a suitable resource state and using measurement-based techniques (see for example Table II of~\cite{Sabapathy2018}). We use the teleportation technique of GKP~\cite{Gottesman2001a}, but this can be translated into an adaptive gate teleportation presented using different gates and additional auxiliary squeezed  states~\cite{Miyata2016a}. We take $\hbar=2$ for the rest of the article. 

For the cubic phase gate the resource state is the cubic phase state defined as $\textsf{V}(\gamma) |0\rangle_p$, which is nonphysical due to the zero momentum ket $\ket{0}_p$. Therefore, as an approximation, we consider $\textsf{V}(\gamma) \textsf{S}(r)^{\dag}\ket{0}$, where $\textsf{S}(r)$ is the standard single-mode squeezing gate given by $\textsf{S}(r)= \exp[r (\hat{a}^2 - \hat{a}^{\dag^2})/2] $. For large squeezing, this state has a large Fock support and hence would be difficult to synthesize directly. To get around this we commute the squeeze operator across the cubic phase gate to obtain $\textsf{S}(r)^{\dag} \textsf{V}(\gamma^{\prime}) \ket{0}$. Let us further assume that $\gamma^{\prime} <<1$, then we can expand the cubic phase gate to first-order in gate strength to obtain $\textsf{S}(r)^{\dag} [1+i\gamma^{\prime} \hat{x}^3/2 ] |0\rangle$. Assuming  that we can apply the on-line squeezing gate using methods such as measurement-based squeezing ~\cite{Miwa2009,Miyata2014,Yoshikawa2008}, we call the remaining terms the resource state~\cite{Marek2011,Yukawa2013a} and it can be expanded in the Fock basis as 
\begin{figure}
\begin{center}
\includegraphics[width=\columnwidth]{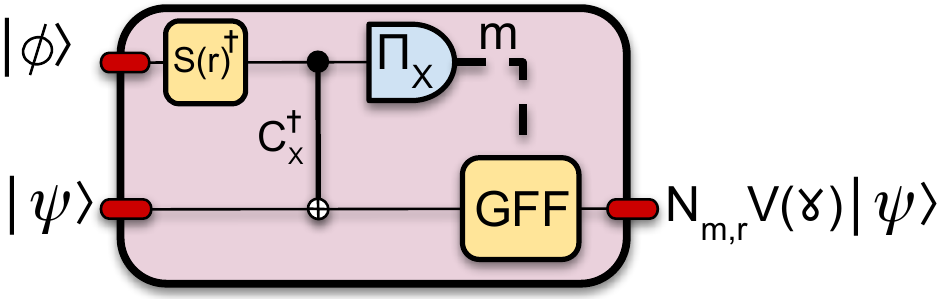}
\end{center}
\caption{The GKP teleportation for implementation of a cubic phase gate~\cite{Gottesman2001a}. Here a state $\ket{\psi}$ is input to one mode along with a resource state $\ket{\phi}$ in the other. $\textsf{S}(r)^{\dag}$ denotes the squeezing operator, $\mathsf{C_x}$ the controlled-X gate given by $\exp [-i \hat{x}_1 \hat{p}_2/2]$, $\Pi_{\rm x}$  the x-homodyne measurement with outcome labeled $m$, \textsf{GFF}  the Gaussian feed-forward correction operator that needs to be applied, $\textsf{N}_{m,r}$ the noise operator, and $\textsf{V}(\gamma)$  the final cubic phase gate that is applied to the input state as shown in Eq.~\eqref{eq:gkp}.   }
\label{fig1:gkp}
\end{figure}
\begin{align}
    \ket{\phi} &= \frac{1}{\sqrt{1+5|a|^2/2}} \left[ \ket{0}  + i a \sqrt{\frac{3}{2}} \ket{1} + i a \ket{3} \right],
    \label{psitar}
\end{align}
where $a \in \mathbb{R}$. 

We now use this resource state in a GKP teleporation scheme as depicted in Fig.~\ref{fig1:gkp}.
The wavefunction of the squeezed resource state is 
\begin{align}
    \tilde{\phi}(x) &= \bra{x} \textsf{S}(r)^{\dag} \ket{\phi} = 
    \int dx' \phi(x') \bra{x} \textsf{S}(r)^{\dag} \ket{x'}.
\end{align}
We note that $\bra{x} \textsf{S}(r)^{\dag} \ket{x'} = e^{r/2}  \langle x | e^r x'\rangle$. So we have that $ \tilde{\phi}(x)= e^{-r} \int dy \phi(e^{-r} y) e^{r/2} \delta(x-y) = e^{-r/2} \phi(e^{-r} x)$.
Then the output state can be derived (similar to Eq.~8 of Ref.~\cite{Sabapathy2018}) as 
\begin{align}
\label{eq3}
    |\psi_{\rm out} \rangle  
    & = N' \exp\left[-\frac{(\hat{x}+m)^2}{4 e^{2r}}\right] \left[1 +  i \frac{\gamma}{2}  (\hat{x}+m)^3 \right] |\psi_{\rm in}\rangle,  \nonumber\\
    \gamma &=  2a e^{-3r}/ \sqrt{6},
        \end{align}
where $N'$ is the normalization factor, $m$ the homodyne measurement outcome, and  $\mathsf{C_x} = \exp [-i \hat{x}_1 \hat{p}_2/2]$ is an entangling gate that can be implemented~\cite{Filip2005,Yoshikawa2008,Ukai2011,Ukai2011a,yokoyama2015}. We now assume that $\gamma \ll1$, allowing us to approximate the terms in the second square bracket as a first-order expansion  in the gate strength of a cubic phase gate, resulting in
\begin{align}
    |\psi_{\rm out} \rangle = N^{'}  \exp\left[-\frac{(\hat{x}+m)^2}{4 e^{2r}}\right] \exp\left[\frac{i\gamma (\hat{x}+m)^3}{2}\right] |\psi_{\rm in}\rangle.  \nonumber
\end{align}
Expanding the terms in the second operator and applying a Gaussian feed-forward  $\mathsf{GFF(m)} = \exp[-i\gamma (3m\hat{x}^2 + 3m^2 \hat{x} +m^3) /2 ]$, we obtain the final action on the input state to be 
\begin{align}\label{eq:gkp}
    |\psi_{\rm out} \rangle = N^{'}  \textsf{N}(m,r) \textsf{V}(\gamma)|\psi_{\rm in}\rangle,
\end{align}
where $\textsf{N}(m,r) = \exp[-(\hat{x}+m)^2/(4 e^{2r})]$ is the Gaussian damping noise operator that depends on the homodyne measurement outcome $m$.

So  using the resource state $\ket{\phi}$, we can effect a transformation which is a weak cubic phase gate along with an unavoidable Gaussian noise factor. Note that the initial squeezing gate $\textsf{S}(r)^{\dag}$ not only reduces the strength of the final cubic phase gate but  also negates the effect of the Gaussian noise operator as seen from Eq.~\eqref{eq3}.  

\begin{figure}
\begin{center}
\scalebox{0.85}{\includegraphics{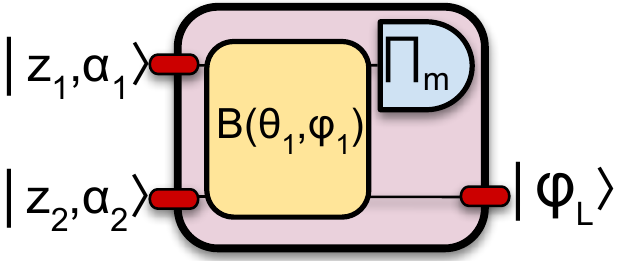}} \vspace{0.25cm}\\
\scalebox{0.7}{\includegraphics{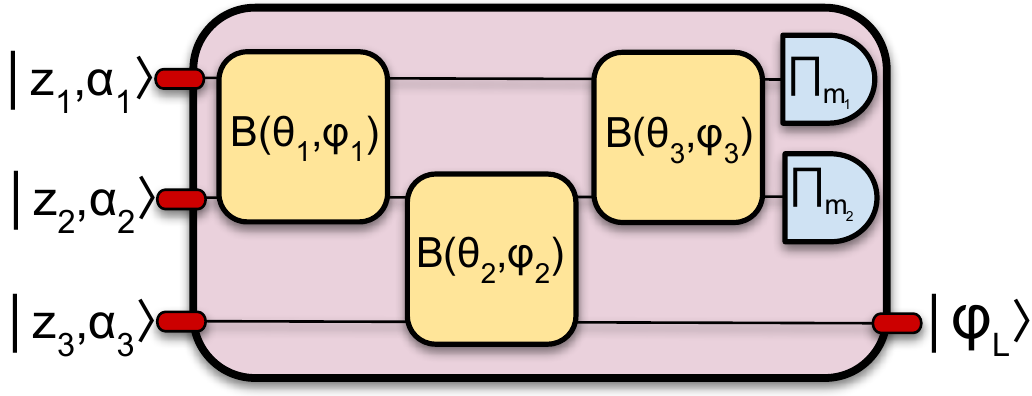}}
\end{center}
\caption{Quantum gadgets of two (top) and three (bottom) modes. $\{z_i,\alpha_i\} \in \mathbb{C} $ denotes the input squeezing $\textsf{S}(z)$ and displacement $\textsf{D}(\alpha)$ gates applied to the vacuum states. $\textsf{B}(\theta,\phi)=\exp[e^{i\phi}a_1a_2^{\dag}-e^{-i\phi}a_1^{\dag}a_2]$ denotes a beamsplitter and $\mathsf{\Pi}_n$ a photon-number resolving detector post-selected to value $n$. $\ket{\phi_L}$ is the output after training the circuit parameters. }
\label{fig2:3mode}
\end{figure}

\section{  Machine learning for state preparation \label{ml}}
The resource state $\ket{\phi}$ can be prepared in the lab using standard sequential photon-subtraction/photon-addition techniques ~\cite{Dakna1999,Dakna1999a,Yukawa2013,Fiurasek2005a,DellAnno2006a}. However, such a method is not scalable since the successful probability of three successive photon additions/subtractions is extremely low, due to the use of very high transmission beamsplitters~\cite{Dakna1999,Gerrits2010} as discussed in Appendix \ref{phosub}. 

To bypass this difficulty, we introduce very different architectures that use Gaussian states conditioned on non-Gaussian  post-selected photon-number resolving (PNR) detectors, akin to Gaussian Boson Sampling circuits~\cite{lund2014boson,hamilton2017gaussian}. Early works have considered conditional measurements on the outputs of beamsplitters for state preparation~\cite{ban1996photon,Dakna1998a,Dakna1998,Dakna1997,clausen1999conditional,Paris2003,knott2016search}. 

\begin{algorithm}[H]
	\caption{ \textsf{StatePrep\_3mode}}\label{alg: main}
	\begin{algorithmic}[0]
		\Function{loss}{$\mb{x}, \textbf{para}, \text{handle}$}:
		\State \# $\mb{x}=\{ \mb{x}_s, \mb{x}_d, \mb{x}_\theta\}$ parameters for squeezing, 
		\State \# displacement and beamsplitter array
		\State $\textbf{para}= \{a, m_1, m_2, \text{cutoff\_dim}\}$
		\State $\ket{\psi_{in}} \gets  \textsf{S}(\mb{x}_s)\textsf{D}(\mb{x}_d)\ket{\mb{0}}$ 
		\State $\ket{\psi_{out}} \gets \textsf{U}(\mb{x}_\theta)\ket{\psi_{in}}$
		\State normIn, normOut $\gets \abs{\langle \psi_{in}\ket{\psi_{in}}}, \abs{\langle \psi_{out}\ket{\psi_{out}}}$
		\State penalty = 100$\times\vert$1-normIn$\vert$ + 100$\times\vert$1-normOut$\vert$
		\State $\ket{\psi} \gets \bra{m_1,m_2}\psi_{out}\rangle$ 
		\State $\text{prob} \gets \langle \psi\ket{\psi}$ 
		\State  $\ket{\psi}_L \gets  \ket{\psi}/\text{prob}$
		\State $\text{fid} = \abs{\langle \psi_L \ket{\phi_T}}^2$ 
		\If{\text{handle} == `fid'} 
		\State \Return -fid +  penalty
		\ElsIf{\text{handle} == `fid\_prob'}
		\State \Return -fid + prob + penalty
		\EndIf 
		\EndFunction
		\vspace{0.02cm}
		\Procedure{optimization}{\textbf{para}, {\em niter}}:
		
		\State \# global exploration to optimize fidelity

		\State \#  \texttt{basinhopping} is a global search algorithm
		
		\State \# further optimize the probability by local search 
		\State \textbf{initialize} $\mb{x}_0$\quad\Comment{randomly chosen with proper range} 
		\State $\mb{x}_1\gets$\texttt{basinhopping}($\mb{x}_0$, \texttt{loss}, \textsf{args}=(\textbf{para}, `fid'),{\em niter}) 	
		\State $\mb{x}_2\gets$\texttt{local}\_\texttt{search}(\texttt{loss}, $\mb{x}_1$, \textsf{args}= (\textbf{para}, `fid\_prob'))
		\State \textbf{save} $\mb{x}_2$
		\EndProcedure
	\end{algorithmic}
\end{algorithm}

Our circuits are depicted in Fig.~\ref{fig2:3mode} for both the two-mode and three-mode architectures.  We then use machine learning algorithms and the \textsf{Strawberry Fields} quantum simulator~\cite{killoran2018strawberry} to train these circuits against the target state given in Eq.~\eqref{psitar}, as presented in Algorithm \ref{alg: main}, where $\ket{\phi_T}$ denotes the target state and $\ket{\psi_L}$ denotes the learned or trained state. The algorithm executes a two-step optimization, where the circuit parameters are first trained to maximize the fidelity with the target state using \textsf{basinhopping} which is a global search heuristic. The second step is then to perform a $\mathsf{local\_search}$ starting from the global optimum found by \textsf{basinhopping}, to further increase the probability of producing the trained state.  We choose the cutoff dimension of each mode in Fig.~\ref{fig2:3mode} to be $15$  such that there is a large enough Hilbert space to be explored but without too much overhead. The PNR detectors in the two-mode case is set to $m=2$ and in the three-mode case to $(m_1,m_2) = (1,2)$ as argued in the following subsection. The details of all the numerical techniques that we use is explained in Appendix \ref{supp:num-tech}.  


\begin{figure}
    \centering
    \includegraphics[scale=0.37]{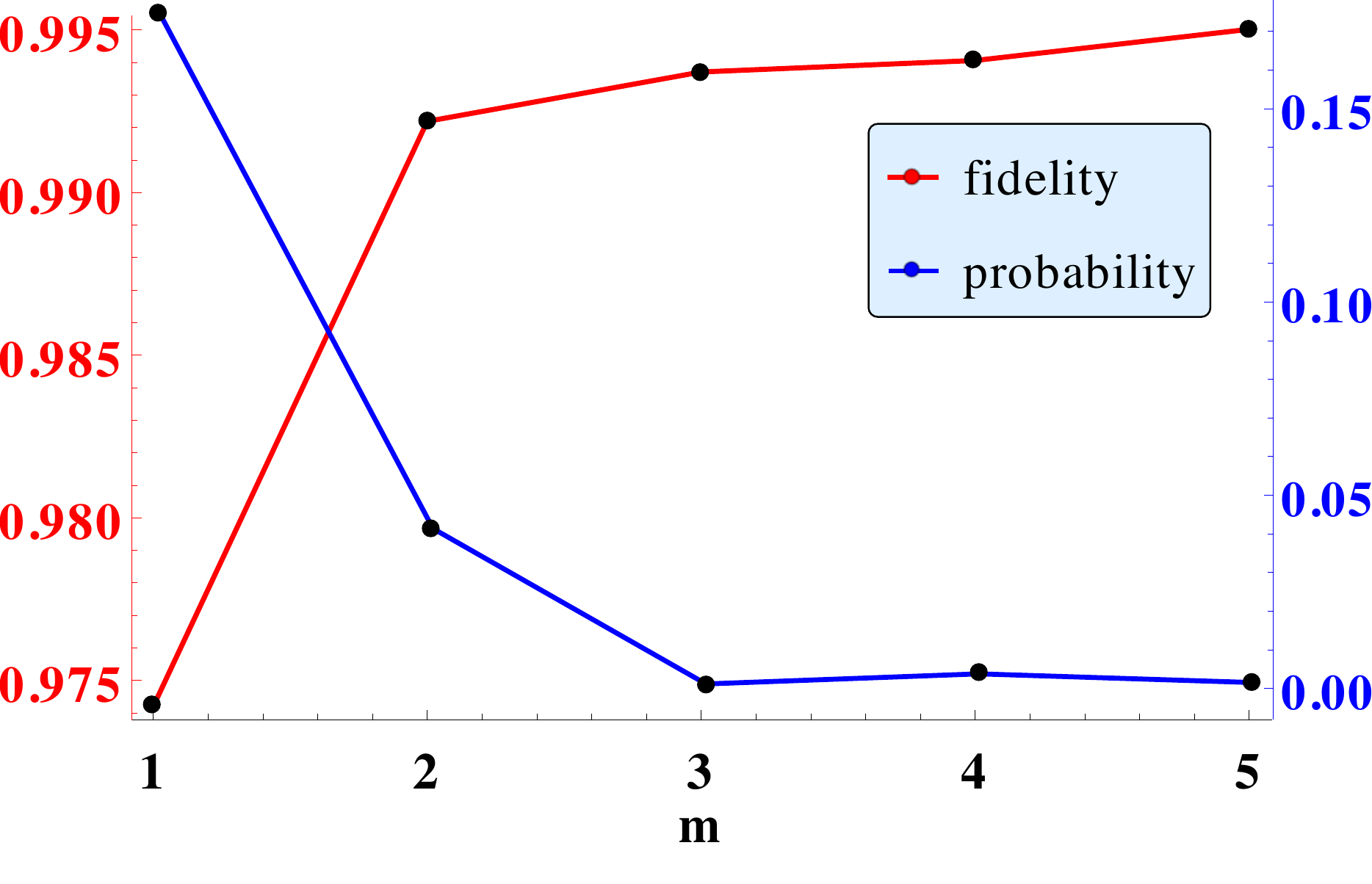}
        \includegraphics[scale=0.32]{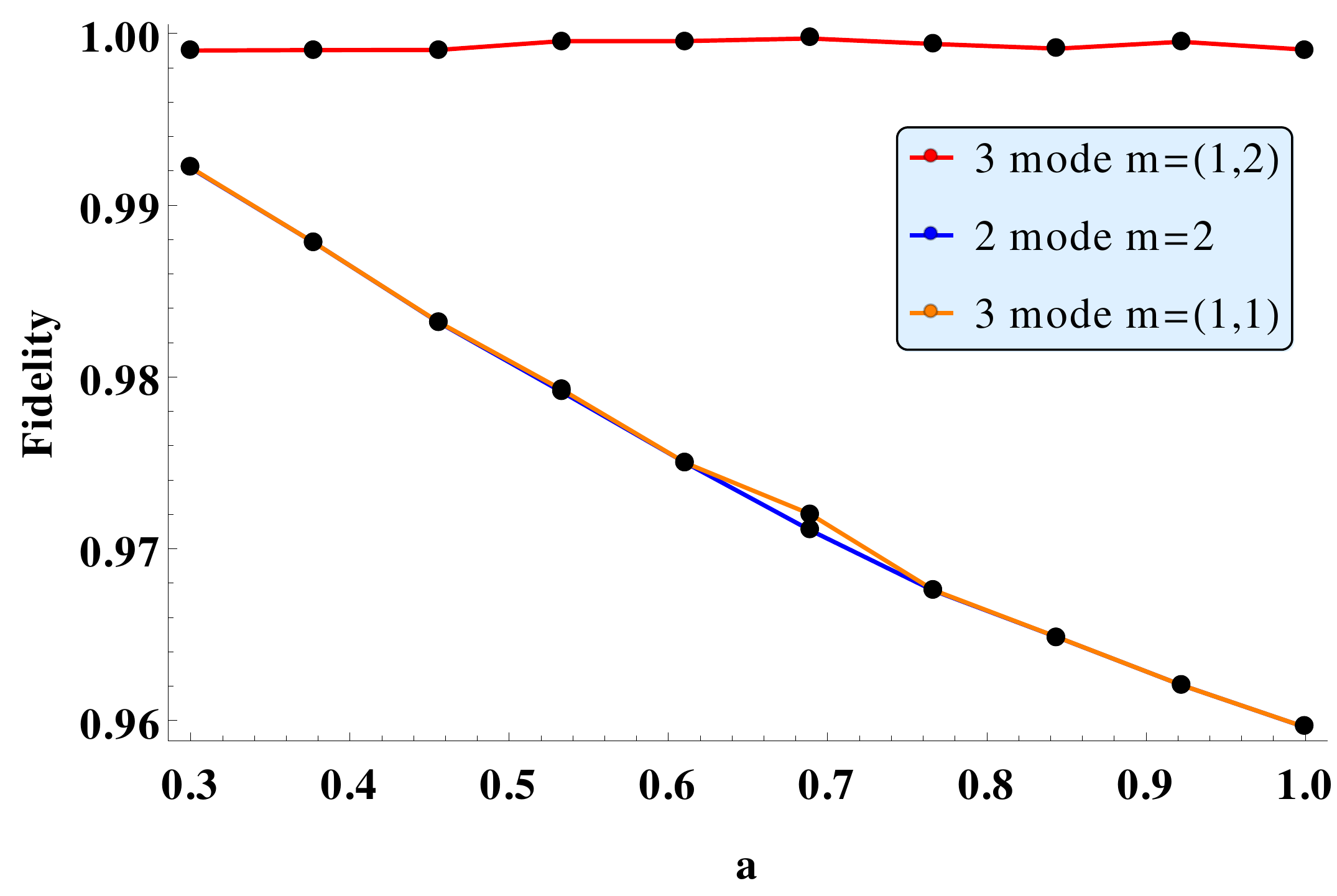}
    \caption{\textbf{Top plot:} Fidelity (rising red line) of the trained state to the target state, and probability (falling blue line) of preparing the trained state as a function of post-selection of the PNR detectors in the two-mode circuit of Fig.~\ref{fig2:3mode}. Here we use \texttt{basinhopping} with {\em niter} = 20 (see App. \ref{supp:num-tech}) without further optimizing the probability. We make the choice of $m=2$ which has a reasonable trade-off between fidelity and probability, both of which we would require to be high. \textbf{Bottom plot:} Comparison of the optimal fidelities for $m=(1,1)$ (bottom-upper orange line) and for $m=(1,2)$ (top red line) for the output state of the three-mode circuit, bench-marked with the two-mode circuit with measurement $m=2$ (bottom-lower blue line), for various values of the target state parameter $a$. The two bottom curves corresponding to the two-mode case with $m=2$ and the three-mode case with $m=(1,1)$ differ only slightly and for $a=0.77$. We find that the measurement setting of $m=(1,1)$ has a very poor performance compared to the one with $m=(1,2)$. 
    For the three-mode case with $m=(1,2)$ and $m=(1,1)$, we use {\em niter} = 40 and {\em niter} = 80 respectively.}
    \label{fig:fid_m}
\end{figure}

\subsection{Post-selection of the PNR detectors \label{supp:pnrfix}}

We provide numerical reasons for our choice of post-selection of the PNR detectors. For the two-mode example we  plot the fidelity and probability as a function of the post-selected PNR measurement in the first plot of Fig.~\ref{fig:fid_m} for a fixed value of $a=0.3$. We find that while the fidelity is increasing with higher photon measurement, the probability drops rapidly. As a reasonable trade-off we pick $m=2$ which gives an increase with respect to the fidelity with a slight cost to the probability of producing the state.

In the second plot of Fig.~\ref{fig:fid_m} we compare the fidelity with respect to the target state parameter $a$ for three settings, namely, (i) two-mode architecture with $m=2$, (ii) three-mode architecture with $(m_1,m_2) = (1,1)$, and (iii) three-mode architecture with $(m_1,m_2)= (1,2)$. 

There are two important findings  that we wish to highlight. The first is that we achieve near-perfect fidelity for the three mode circuit for all values of $a \in [0.3,1]$ when the PNR detectors are post-selected to values $(m_1,m_2)=(1,2)$. The second interesting numerical observation is that the performance of the two-mode architecture with $m=2$ is extremely close to the three-mode circuit with $(m_1,m_2) = (1,1)$. In both cases a common feature is that the sum of the values of post-selected PNR detectors are the same. We anticipate that a deeper understanding is possible if we explore these aspects from the perspective of the resource theory of non-Gaussianity ~\cite{pang,quntao18,Takagi2018,albarelli2018resource,lami2018all}.

\subsection{Trained parameters for two-mode and three-mode circuits}

\begin{table*}
    \centering
\begin{filecontents*}{2mode.csv}
$a$,{\color{red}$r_1$},{\color{red}$r_2$},$\phi^r_1$,$\phi^r_2$,$d_1$,$d_2$,$\phi^d_1$,$\phi^d_2$,$\theta$,$\psi$
0.3,{\color{red}0.37},{\color{red}-0.37},-3.08,0.59,-0.25,0.35,0.77,-0.98,-0.64,2.26
0.38,{\color{red}-0.43},{\color{red}-0.43},-1.72,-1.64,-0.36,-0.3,-2.05,-3.25,2.25,-1.93
0.46,{\color{red}-0.48},{\color{red}-0.48},-1.34,0.59,0.34,-0.39,-0.14,-0.93,2.49,2.85
0.53,{\color{red}-0.5},{\color{red}0.46},1.73,0.37,0.4,-0.29,-0.64,-1.2,0.97,-0.77
0.61,{\color{red}-0.54},{\color{red}-0.58},0.55,0.65,0.39,-0.41,0.75,2.28,-0.68,-1.26
0.67,{\color{red}-0.34},{\color{red}-0.48},3.1,0,-0.14,0.39,1.55,-1.57,0.61,-3.12
0.77,{\color{red}0.35},{\color{red}-0.51},0.85,0.01,0.14,-0.4,-1.14,1.57,0.63,2.72
0.84,{\color{red}0.28},{\color{red}-0.46},-0.57,0,0.07,0.38,4.42,-1.57,0.62,-2.86
0.92,{\color{red}-0.53},{\color{red}0.34},-1.84,0.02,-0.4,0.12,-2.49,7.9,-0.93,0.93
1,{\color{red}-0.23},{\color{red}0.43},-0.97,3.14,0.01,-0.38,-0.47,-1.57,3.76,-1.09
\end{filecontents*}
\DTLloaddb[keys={1,2,3,4,5,6,7,8,9,10,11}]{mytable2}{2mode.csv}
\renewcommand{\dtldisplayafterhead}{\hline \hline}
\DTLdisplaydb{mytable2}
    \caption{Optimal circuit parameters for the two-mode architecture with PNR $m=2$ (top schematic of Fig. \ref{fig2:3mode}). The state parameter $a$ of Eq. \eqref{psitar} is varied in equal steps in the range $[0.3,1]$.  $\{r_i,\phi^r_i\}$ (\{$d_i, \phi^d_i$\}) are the magnitude and phase for squeezing (displacement) applied to the $i$-th vacuum mode.  $\{\theta,\phi\}$ are the parameters for the beamsplitter. }
    \label{tab:2mode}
\end{table*}


\begin{table*}
    \centering
\begin{filecontents*}{3mode.csv}
$a$,{\color{red}$r_1$},{\color{red}$r_2$},{\color{red}$r_3$},$\phi^r_1$,$\phi^r_2$,$\phi^r_3$,$d_1$,$d_2$,$d_3$,$\theta_1$,$\theta_2$,$\theta_3$,$\phi_1$,$\phi_2$,$\phi_3$
0.3,{\color{red}-0.27},{\color{red}0.65},{\color{red}0.66},-4.14,0.53,-1.94,0.38,-0.19,-0.47,2.29,-4.19,3.5,2.77,0.59,-0.2
0.38,{\color{red}-0.53},{\color{red}-0.75},{\color{red}-0.55},7.19,10.16,10.73,0.51,-0.03,0.53,2.3,-2.04,-4.11,-0.29,-1.99,-0.89
0.46,{\color{red}-0.75},{\color{red}-0.54},{\color{red}0.27},1.23,-5.31,1.53,-0.04,0.38,0.63,0.7,1.97,-0.88,-1.62,0.72,-1.79
0.53,{\color{red}0.71},{\color{red}0.67},{\color{red}-0.42},-2.07,0.06,-3.79,-0.02,0.34,0.02,-1.57,0.68,2.5,0.53,-4.51,0.72
0.61,{\color{red}0.72},{\color{red}0.65},{\color{red}-0.45},0.23,0.49,-3.81,0,-0.33,-0.01,-1.57,-2.46,0.63,-1.81,-0.22,6.42
0.67,{\color{red}0.52},{\color{red}0.56},{\color{red}0.74},0.09,3.35,-3.26,0.64,-0.33,0.02,-3.31,2.29,2.63,-3.33,1.41,-6.29
0.77,{\color{red}-0.55},{\color{red}0.73},{\color{red}0.03},1.14,0.98,-3.03,0.34,0.11,0.51,-2.3,-1.96,0.69,-3.22,4.32,-4.07
0.84,{\color{red}0.74},{\color{red}0.54},{\color{red}-0.48},0.5,-5.29,2.73,-0.01,-0.36,0.01,-1.52,0.7,-3.8,-0.95,-1.06,-1.24
0.92,{\color{red}-0.54},{\color{red}0.49},{\color{red}0.72},-0.44,2.36,3.14,0.15,-0.48,0.25,-1.3,2.27,0.67,-2.13,3.94,1.99
1,{\color{red}0.66},{\color{red}-0.38},{\color{red}-0.76},3.36,-0.73,0.4,-0.05,-0.82,0,1.56,-2.32,0.61,-3.67,-0.97,-0.71
\end{filecontents*}
\DTLloaddb[keys={1,2,3,4,5,6,7,8,9,10,11,12,13,14,15,16}]{mytable}{3mode.csv}
\renewcommand{\dtldisplayafterhead}{\hline \hline}
\DTLdisplaydb{mytable}
    \caption{Optimal circuit parameters for the three-mode architecture with PNR $(m_1=1,m_2=2$) in the bottom schematic of Fig. \ref{fig2:3mode}.  $\{r_i,\phi^r_i\}$ are the magnitude and phase for squeezing applied to the $i$-th vacuum mode. $d_i$  are the real displacements applied to the $i$-th vacuum mode. It turns out that taking the displacements to be real gives rise to more stable solutions. $\{\theta_i,\phi_i\}$ correspond to the parameters for the $i$-th beamsplitter. }
    \label{tab:3mode}
\end{table*}

The final optimization values for the two-mode circuit parameters and the three-mode circuit parameters are listed in Tables \ref{tab:2mode} and \ref{tab:3mode}, respectively. We find that the required squeezing values $r$  are all $\leq 5.1$ dB for the two-mode case and  $\leq 6.7$ dB for the three-mode case, which is a very experimentally accessible value. Further, unlike the two-mode case, it turns out that taking the displacements to be real provided better fidelity results for the three-mode case. 
We wish to highlight that the number of steps for which we run the optimization algorithms are fixed for all values of state parameter $a$ for the sake of reproducibility of the numerical results. For certain values of $a$ an improvement in probability cannot be ruled out.

\begin{figure*}
    \centering
    \scalebox{0.32}{
    \includegraphics{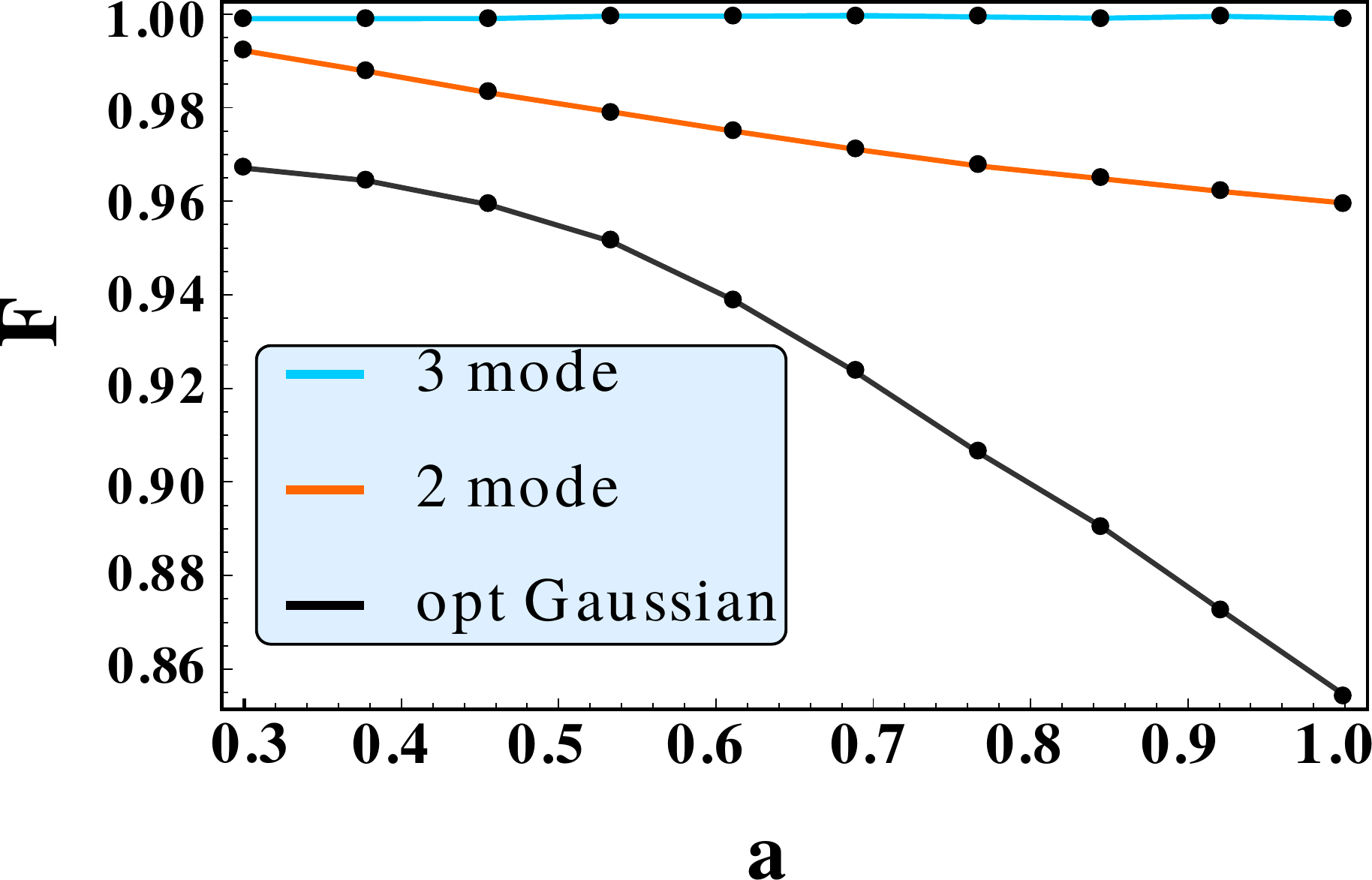}~~\includegraphics{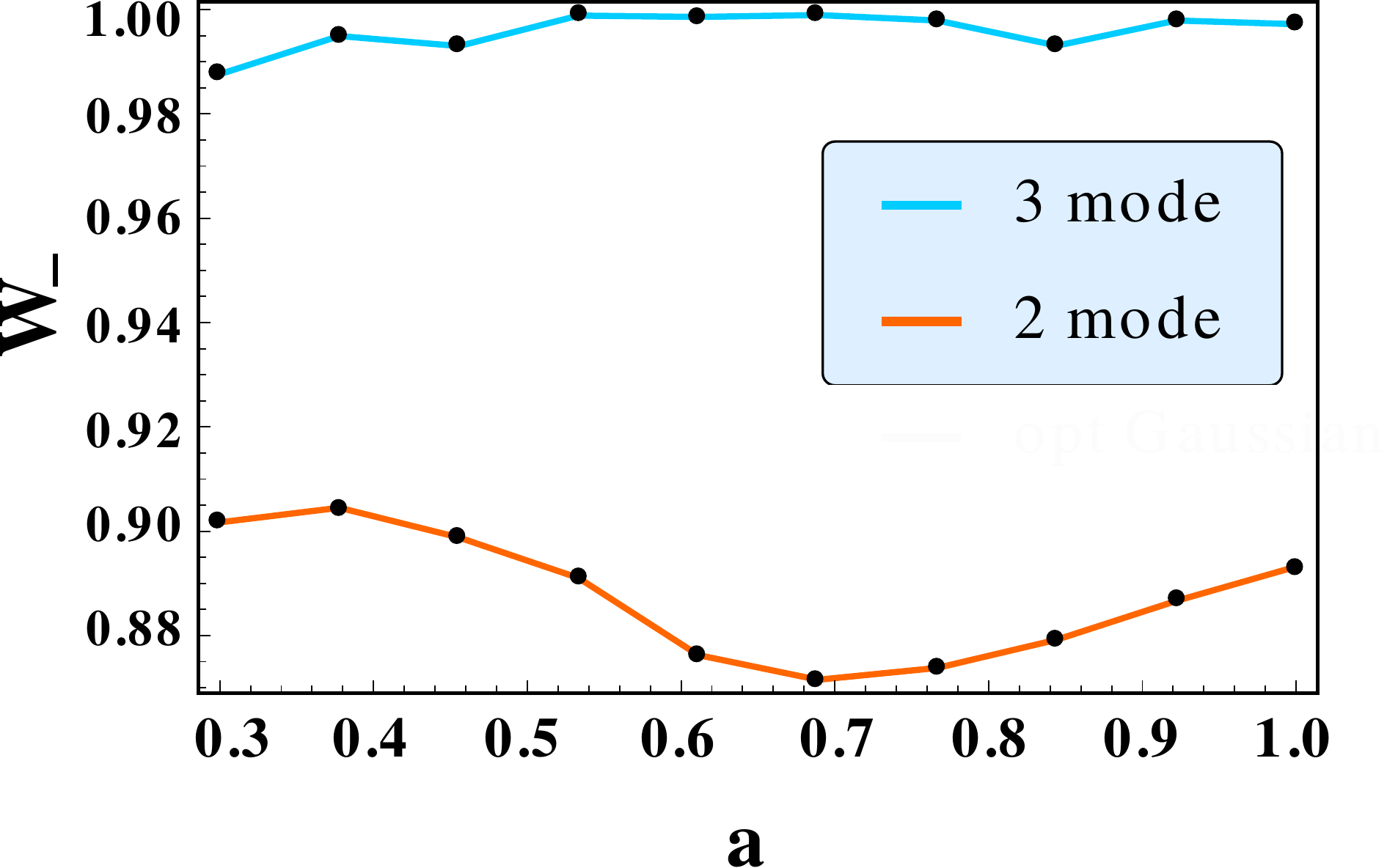}\hspace{-0.9cm} \includegraphics{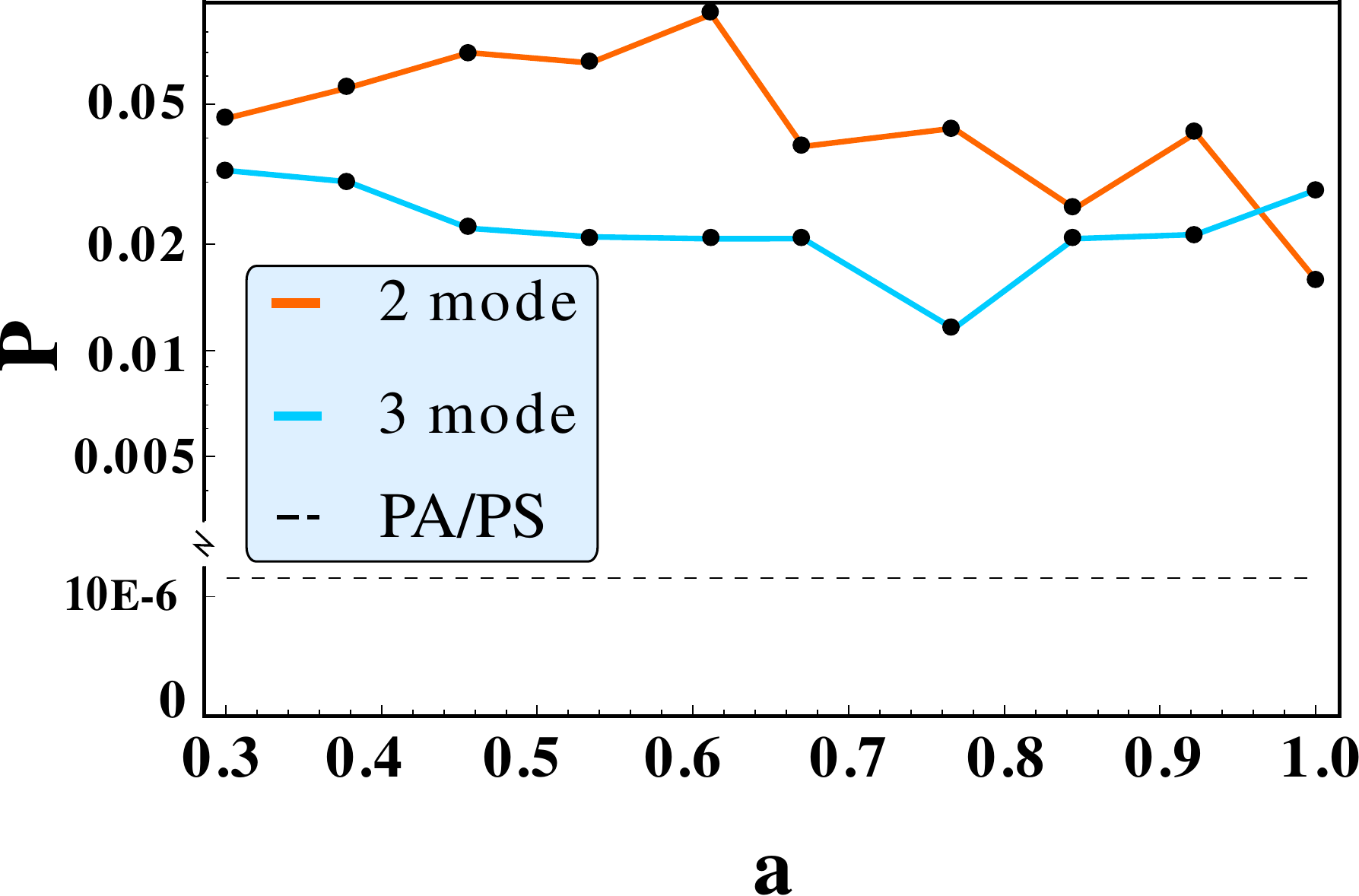}}
    \caption{Left plot depicts the optimal fidelity $F$ of the machine learned state $\ket{\phi_L}$ in three-mode (top blue line) and two-mode (middle red line) architectures, and the closest Gaussian state (bottom black line) with respect to the target state $\ket{\phi}$ in Eq.~\eqref{psitar}. Middle plot shows the overlap of the negative region of the Wigner function $W_-$ between $\ket{\phi_L}$ and $\ket{\phi}$ for the three-mode (top blue line) and the two-mode (bottom orange line) cases.  Right plot depicts the probability $P$ of the optimal output state for the two-mode (top orange line) and three-mode (middle blue line) architectures  compared with three consecutive photon-additions/subtractions (PA/PS) with a probability  $\sim 10^{-6}$ (bottom dashed line).  We see that for fidelity and Wigner negativity, the three-mode case performs best with near-perfect fidelity at the cost of a drop in probability when compared with the two-mode case.  }
    \label{fig:trained}
\end{figure*}

\subsection{Physical comparison of the two-mode and three mode cases}
 The trained and target states are both pure, and we have that the fidelity  is equivalent to the Wigner overlap~\cite{cahill1969}, i.e., 
\begin{align}
    F(\phi_1, \phi_2) &= |\braket{\phi_1|\phi_2}|^2 = \int dx dp \, W(x,p;\phi_1) W(x,p;\phi_2). \nonumber \end{align}
Since the negativities of the Wigner function are crucial, we define by $W_{-}$ the overlap of the negative region of the Wigner functions of the  output and  target state, each negative region being renormalized to $1$. For parameter $a \in [0.3,1]$ we have that the gate strength $\gamma \in [0.0122,0.0407]$ (by Eq. \eqref{eq3}) if the initial  squeezing gate in Fig. \ref{fig1:gkp} had $r=1$.

We now use the optimization values of the two-mode and three-mode cases in Tables \ref{tab:2mode} and \ref{tab:3mode} to plot the fidelity, the overlap of the negative region of the Wigner function between  the trained state $\ket{\phi_L}$ and the target state $\ket{\phi}$, and the probability of producing the trained state in Fig.~\ref{fig:trained}.

\begin{figure*}
    \centering
       \scalebox{0.3}{\includegraphics{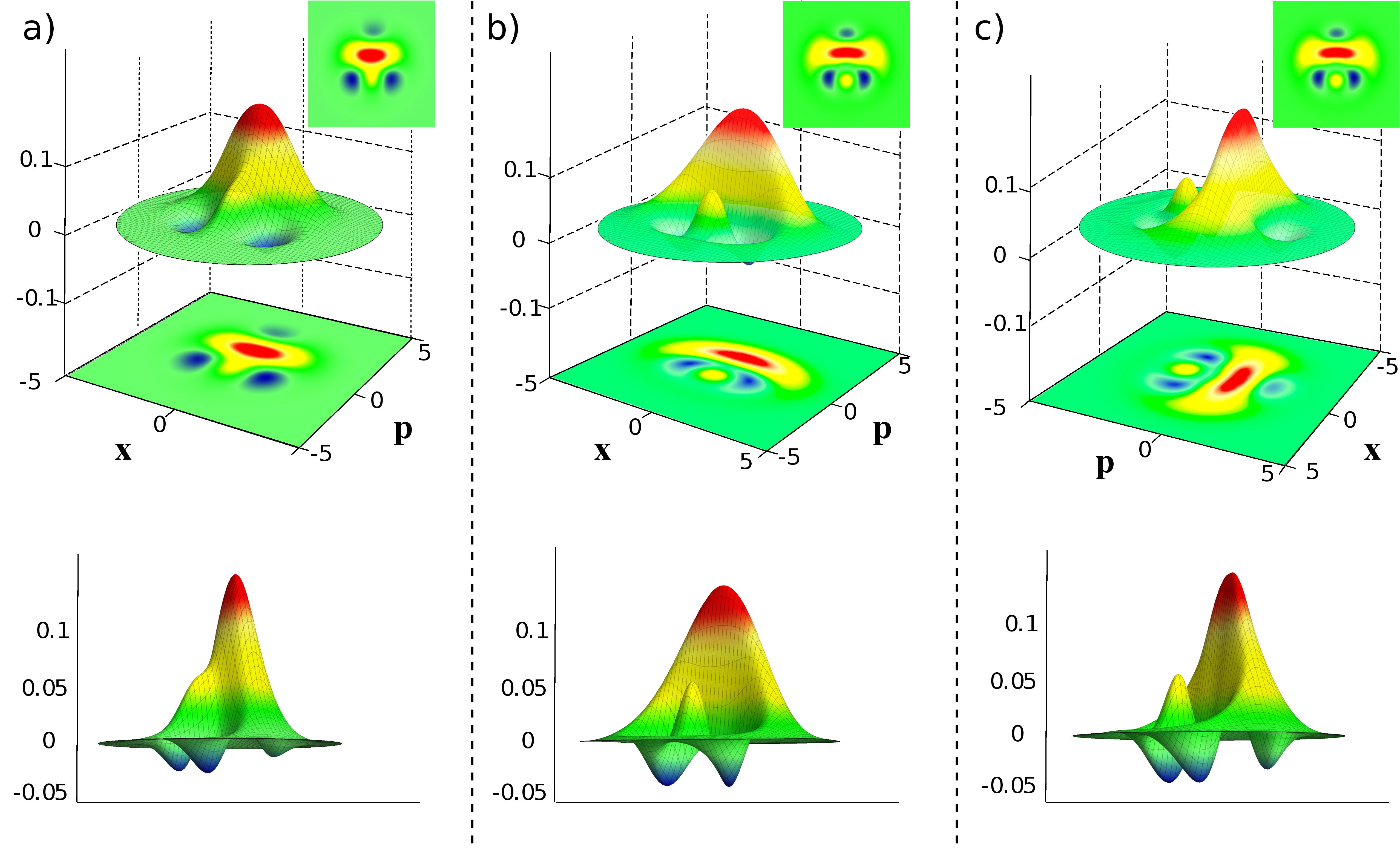}}
    \caption{The Wigner function of the trained state for three cases (a) $a=0.3$ with three-mode architecture, (b) $a=0.61$ with two-mode circuit, and (c) $a=0.61$ with the three-mode architecture. The insets are the Wigner contour plot for the target states for the corresponding values of $a$. We see that the trained state using the three-mode circuit is indistinguishable with the target state. We find that for the two-mode case (b) with $a=0.61$, one Wigner negative region is missing with respect to the three-mode case (c), thereby leading to lesser fidelity to the target state. }
    \label{fig:wig}
\end{figure*}

We find that using the three-mode architecture the trained state is extremely close to the target state, and so is the overlap of the negative region of the Wigner function (middle plot of Fig. \ref{fig:trained}). On the other hand, the two-mode circuit performs better than the three-mode circuit in terms of the probability of producing the trained state, but there is a substantial drop in the corresponding $W_{-}$.

To provide a visual comparison of the two-mode and three-mode performance, we plot the Wigner functions for the trained states for specific values of the target state parameter $a$ in Fig.\,\ref{fig:wig}. In each of the subplot, the insets at the top-right corner are contour plots of the Wigner function of the target state. In subplot (a) we consider the three-mode gadget with $a=0.3$. We find a very good match between the trained state and the target state. In subplots (b) and (c) we consider $a=0.61$, where (b) corresponds to the two-mode gadget and (c) to the three-mode gadget. We find an important difference that while the three-mode gadget produces a trained state that is close to the target state, the two-mode gadget produces a state with one of the Wigner negative `dips' missing. This leads to a reduced target state fidelity as already mentioned in Fig. \ref{fig:trained}.



\subsection{Generating random states}
\begin{figure}
\begin{center}
\scalebox{0.35}{\includegraphics{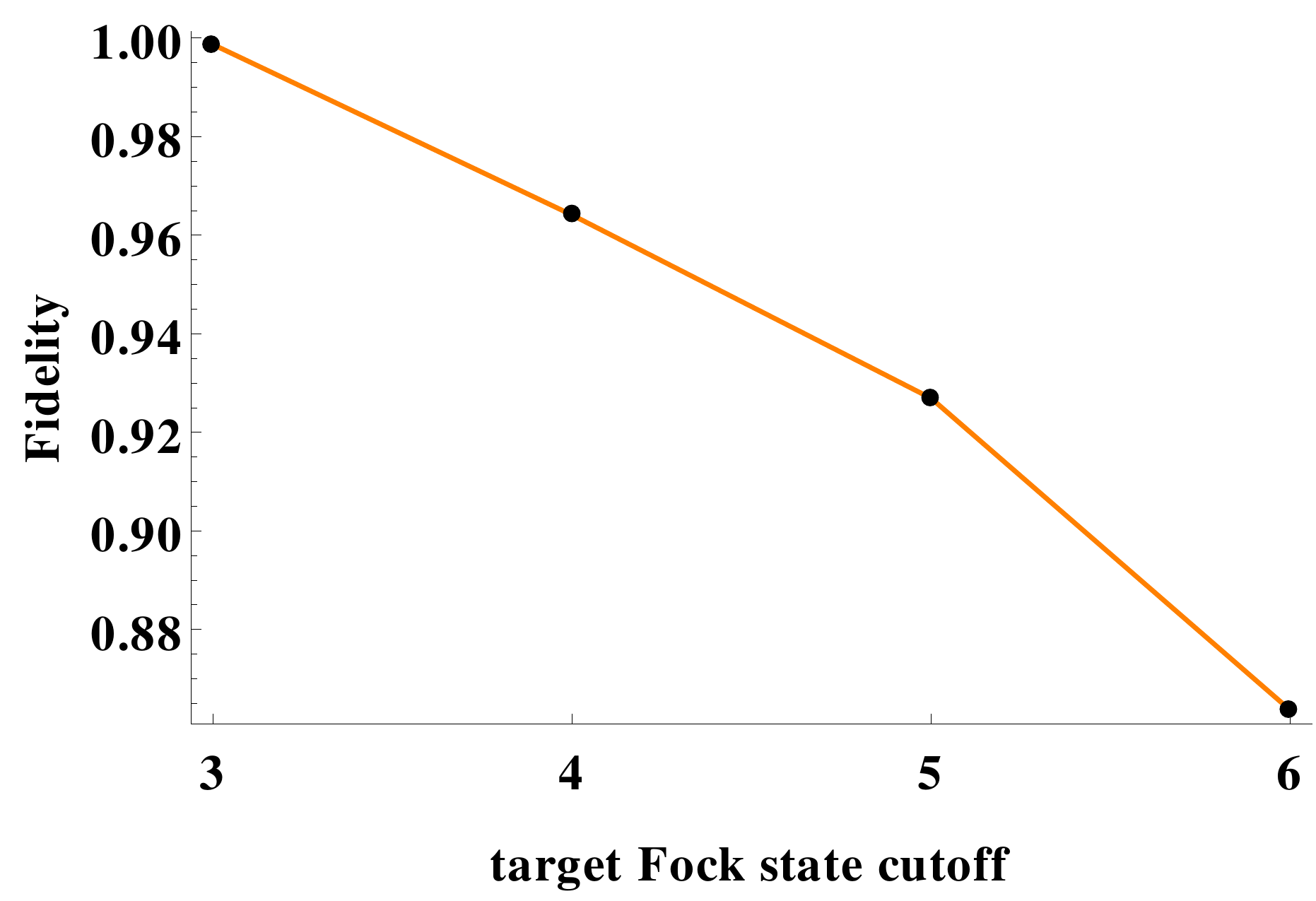}}
\end{center}
\caption{Plot of the average fidelity of the trained state  to a target random state $\ket{\phi_T}$ for various values of the Fock cutoff dimension $n_c$, using the three-mode gadget with $(m_1,m_2) = (1,2)$. We sampled 100 random states for each value of target cutoff dimension. We find that using the three-mode gadget any random superposition of up to three photon Fock states can be generated with near-perfect fidelity for this choice of post-selected PNR detection pattern. }
\label{random_state}
\end{figure}

As a final numerical experiment we try to target a random state for each value of Fock state cutoff as outputs of the three-mode architecture  with the same post-selected PNR detectors $(m_1,m_2) = (1,2)$. The target state is now of the form $\ket{\phi_T}= \sum_{n=0}^{n_c} c_n \ket{n}$, where $n_c$ is the maximum Fock support or cutoff, and $\{c_n \}$'s are randomly chosen coefficients that are normalized to unity. In Fig.~\ref{random_state} we plot the average fidelity of the trained state to the target random state for various values of the cutoff dimension of the random state. As expected the average fidelity is monotonically decreasing with an increase in the output cutoff dimension of the target state. However, for cutoff Fock state value $n=3$ we find that the average fidelity is extremely close to 1.

\section{ Quantum resource farms to improve success probability \label{qrf}}
\begin{figure}
\begin{center}
\scalebox{0.7}{\includegraphics[width=\columnwidth]{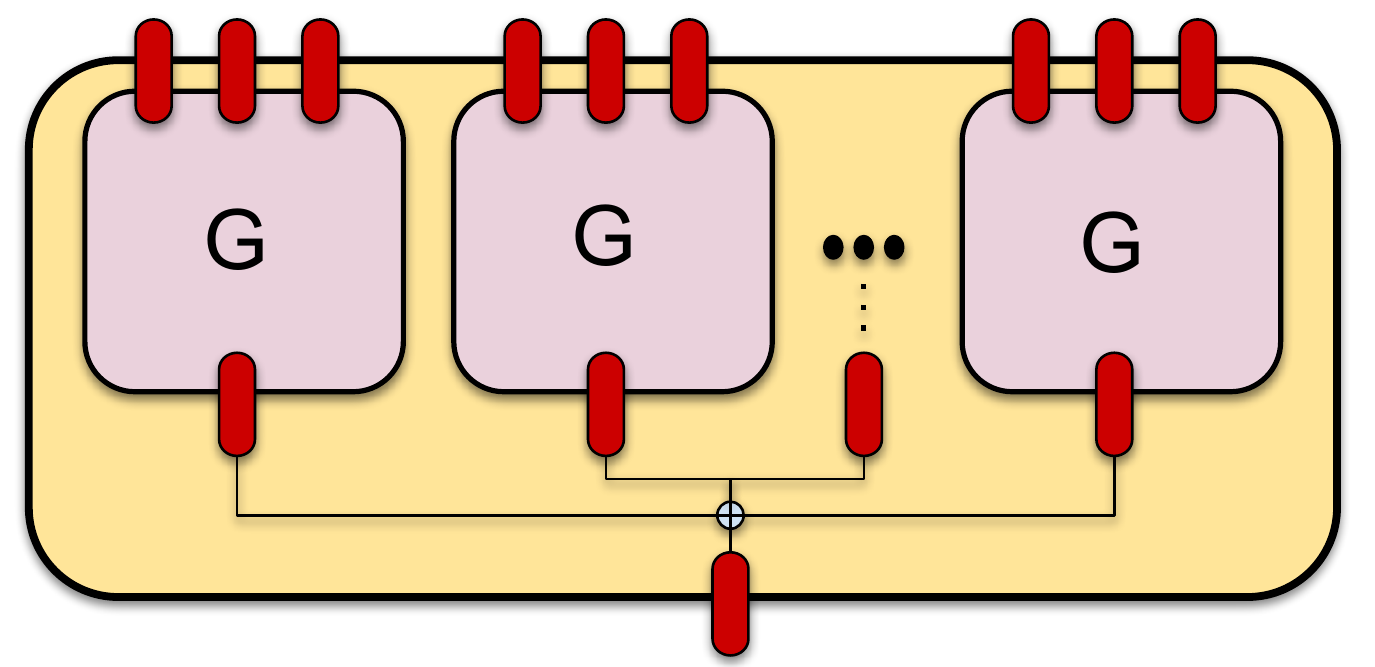}}
\end{center}
\caption{A quantum resource farm that consists of copies of quantum gadgets \textsf{G} of Fig.~\ref{fig2:3mode} (three-mode case shown here) placed in a parallel manner where the individual outputs are collected to one output. There is a trade-off between the rate of production of the state and the number of gadgets used. }
\label{fig:rf}
\end{figure}

The resource state preparation is probabilistic since it is conditioned on a particular post-selected measurement outcome as shown in Fig.~\ref{fig2:3mode}. To improve the success probability of the resource state preparation, we propose an optical setup which we call a quantum resource farm. We line up identical copies of the state preparation gadget in parallel, and connect all outputs to a single wire or sink as shown in Fig.~\ref{fig:rf}. We then find a trade-off between the success probability and the number of gadgets in the resource farm, when allowing for a small  error.

\begin{figure}
\begin{center}
\scalebox{0.65}{
\includegraphics[width=\columnwidth]{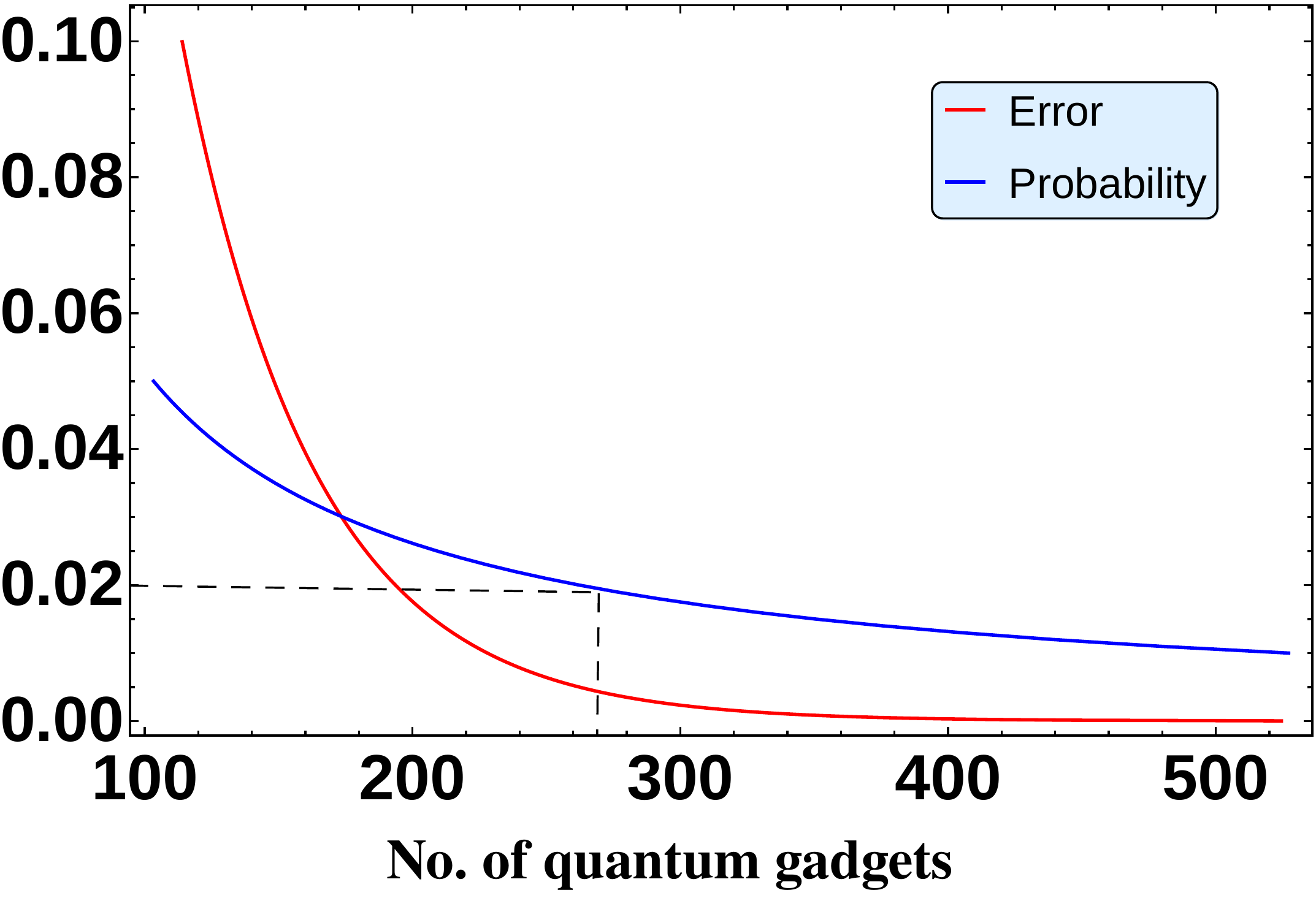}}
\end{center}
\caption{The error (top red line) for a probability fixed at $0.02$ and probability (bottom blue line) for error fixed at $0.005$ as a function of the number of quantum gadgets in the resource farm. As an example, if we consider $p=0.02$ and $\epsilon =0.005$, i.e., the  gadget produces the resource state with efficiency of $2\%$ and if we want the farm to have a success rate of $99.5\%$, then we require $\sim 260$  gadgets (dotted lines) in the farm. }
\label{fig:foo}
\end{figure}

The model of our resource farm has $n$ parallel quantum gadgets, and let $p$ be the probability of success of producing the resource state from each one of them. Let $\epsilon$ be the error which captures the degree of determinism of the entire farm. The probability of producing the required resource state from the farm is denoted $P(F)$ and is given by the union of events of any of the constituent gadgets preparing this state. Even in the case where multiple resource states are simultaneously prepared, we continue to count it as a useful event. Thus, this success probability is identical to the complement event probability that none of the gadgets produce the required resource state;
\begin{align}\label{succp}
    P(F) = 1- (1-p)^n.
\end{align}
Allowing an error $\epsilon$ for the failure of the state preparation from the farm, we set 
    $P(F) \geq 1- \epsilon$.
Therefore, the minimum required number of gadgets $n_{\rm min}$ is given by
\begin{align}\label{nmin}
    n_{\rm min} = \left \lceil \frac{\log \epsilon}{ \log(1-p) } \right \rceil,
\end{align}
where $\lceil y \rceil $ denotes the smallest integer $\geq y$. If we assume an ideal GKP-teleportation using this resource state, then the effective cubic phase gate can be implemented at the same rate as the resource state preparation. If we use a fewer number of gadgets than $n_{\rm min}$, the rate at which the cubic phase gate can be implemented will be proportionally reduced.

\subsubsection*{Examples} The top (red) line in Fig.~\ref{fig:foo} shows the variation of error $\epsilon$  as a function of the number of quantum gadgets $n_{\rm min}$ for a fixed probability of $p=0.02$ of state preparation from a  gadget. The bottom (blue) line shows the variation of $p$ vs $n_{\rm min}$ for $\epsilon = 5\cdot 10^{-3}$. For $p =0.02$ (i.e., $2\%$ efficiency) and  $P(F)= 0.995$ ($\epsilon = 0.005$), $n_{\rm min} \sim 260$ (shown by dotted lines).  

So we that there is a trade-off between the number of gadgets used and the success probability as captured by Eq. \eqref{succp}. If one requires near-determinism using this method to increase success probability, we see that the resource cost is  challenging for the example mentioned in the previous paragraph. However, the analysis provides us with a way to estimate the resources required to increase the success probability. 

\section{Noise analysis in a quantum gadget \label{noise}}
\begin{figure}
\begin{center}
\includegraphics[width=\columnwidth]{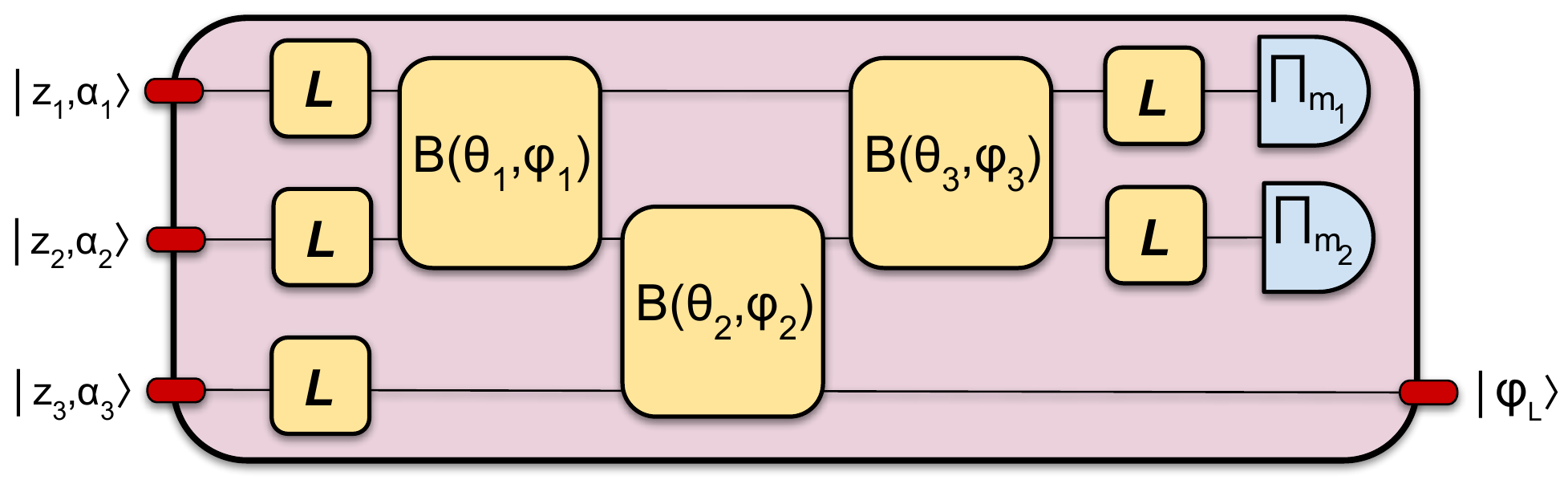}
\end{center}
\caption{Addition of source and detection losses to the three-mode gadget presented in Fig. \ref{fig2:3mode}. The loss is denoted by $L$ and is modeled by a pure-loss bosonic channel $L(\eta)$ where $\eta \in [0,1]$ is the transmission coefficient. }
\label{3mode-loss}
\end{figure}
In the ideal case the state in Eq. \ref{psitar} is produced with near-perfect fidelity and a probability $> 1\%$ as shown earlier in Fig. \ref{fig:trained}. However, in practical scenarios there are various sources of imperfections in the optical implementation. Of these, we focus on the state preparation stage where sqeeuzed vacuua are produced and in the mearsurement stage where a subset of modes are subjected to a photon number-resolving detection. We model these imperfections by a photon-loss channel as described in the following section.

We now consider in Fig.~\ref{3mode-loss} a noisy version of the 3-mode gadget used to produce weak cubic phase states. The source and measurement losses are modelled using the pure-loss channel denoted by $\textsf{L}(\eta)$ with transmission coefficient $\eta \in [0,1]$, a completely positive trace preserving bosonic Gaussian channel with well known Kraus operators \cite{ivan2010,albert2018performance} 
\begin{align}
\label{lossop}
    \textsf{A}_k(\eta) &= \left(\frac{1-\eta}{\eta}\right)^{k/2} \frac{a^k}{\sqrt{n!}} (\sqrt{\eta})^{a^{\dag}a}, \nonumber\\ 
    \textsf{L}(\eta)[\rho] &=  \sum_{k=0}^{\infty} \textsf{A}_k(\eta) \rho \textsf{A}_k(\eta)^{\dag}.
\end{align}
For simplicity we fix the detection efficiency to $96\%$ and we take the noise at the source to be identical in the three modes. We vary this noise and consider the corresponding relation to properties of the output state. 

\begin{figure}
\begin{center}
\scalebox{0.4}{\includegraphics{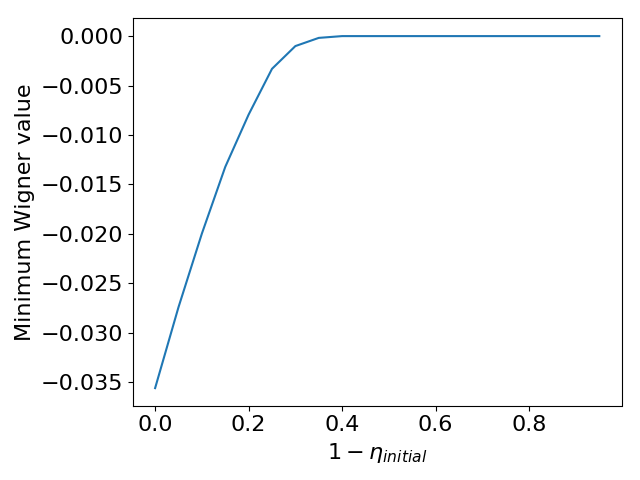}} \\ \scalebox{0.4}{\includegraphics{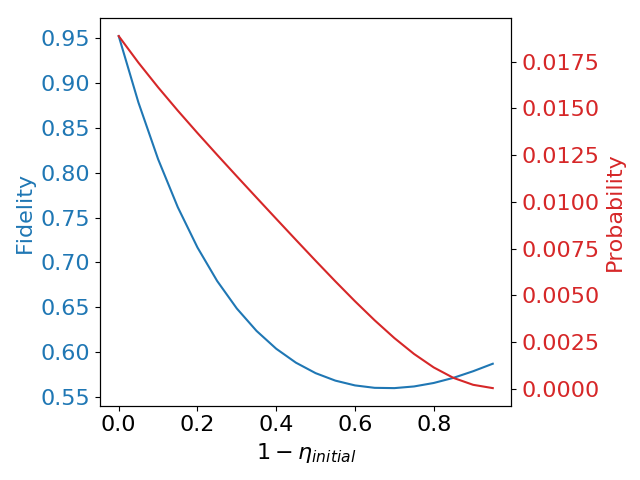}}
\end{center}
\caption{The effect of noise on the preparation of target states in Eq. \eqref{psitar} using the optical circuit of Fig.~\ref{3mode-loss}. The top figure depicts the variation of the minimum value of the Wigner function of the output state as a function of the circuit loss $\ell = 1 - \eta_{\rm initial}$.  Here we set the measurement loss to a fixed value of $4\%$ and vary only the preparation loss parameter. We find that for the output state parameter $a=0.3$, the amount of loss the circuit can tolerate before all the Wigner negativities are washed out is approximately $\ell_T =0.4$. The bottom plot shows the variation of fidelity (bottom blue curve) of the output state with regard to the target state along with the probability (top red curve) of producing the state. The values of fidelity and probability monotonically decreases for increase in the noise value up to the threshold value. }
\label{onloss}
\end{figure}

In Fig.~\ref{onloss} we plot the fidelity, probability, and the minimum value of the Wigner function of the output state from the gadget. The fidelity is computed with respect to the target state (Eq. \eqref{psitar})  as a function of source loss where ${\rm loss} = 1- \eta_{\rm initial}$, and $\eta_{\rm initial}$ depicts the transmission at the source. For convenience, we fix the loss at the detection part of the circuit to $4\%$ and the target state parameter $a$ to $0.3$. The other parameters of the circuit are chosen to be the optimal values from the first row in Table \ref{tab:3mode}.

We find that there is a loss threshold $\ell_T$ beyond which all the Wigner negativity of the output state gets washed out. For the value of the target state parameter $a=0.3$, we find that this threshold is at $\ell_T \sim 0.4$, i.e. at $\eta_{\rm initial } = 0.6$. Any value of loss greater than $40\%$ produces output Wigner functions that are positive everywhere. As expected we find that both the fidelity and the probability of producing that state monotonically decrease up to the loss value $\ell_T$. Beyond this value of loss the fidelity does increase a bit, but the dependence of the fidelity and probability are not of particular interest in this region~\cite{mari2012positive} since the Wigner negativities have already been washed out. On the other hand, we also find that if the state preparation fidelity is required to be high $(> 90\%)$, then $\eta_{\rm initial}$ needs to be $>0.9$, i.e. the source loss needs to be well below $0.1$.

\section{Conclusion \label{con}} We proposed a novel method for the production of target resource states using machine learning techniques.  We first constructed quantum gadgets that produce very general Gaussian pure states in two and three-modes which were conditioned on all but one of the modes on a post-selected photon-number resolving detector. We then tuned the parameters of the gadget to produce a trained state that is of almost perfect fidelity with the target state. Fig. \ref{random_state} indicates that this three-mode gadget and particular post-selection could have a more universal property to produce a state with any choice of superposition in the first four Fock basis states. 

We found that our architecture in Fig. \ref{fig2:3mode} led to an increase in the probability (of the order of $10^4$) of producing the target state compared to conventional sequential photon-addition or subtraction methods. 
As a way to increase even this success probability, we proposed the notion of a quantum resource farm that connects such quantum gadgets in parallel. We then obtained the trade-off between the new success probability and the number of required quantum gadgets. It turns out that the resource costs are high to achieve near-determinism using this method as a means to circumvent quantum memories. Also, in this method, the rate of producing the resource states is limited by the measurement rate of the photon number-resolving detectors. 

We expect that our architecture for a quantum gadget can be realized in the  near-term, and we anticipate substantial improvements in future through technological advancements driven by demand. Also the squeezing requirements for our state preparation (Eq.\,\eqref{psitar}) is within reasonable bounds of $<7{\rm dB}$. 
Having fewer gadgets in the resource farm would lead to a lower production rate of these gates while not compromising on the  fidelity.  Also, photon-loss plays an important role in the state preparation and needs to be accounted for as mentioned in Sec. \ref{noise}.  

Our proposal is only an initial step in this direction and there is scope for improvements and optimization using tools such as machine learning. There are two ways to obtain higher gate strengths. One is to concatenate many weak cubic gates considered here, and two, is to consider resource states that have a higher Fock support than those considered here. Our methods could prove useful in both these avenues.  Further, the final applied cubic phase gate using the resource state and gate-teleportation circuit tends to be noisy, and requires additional considerations. 
Finally, the detailed numerical analysis also seems to suggest that the PNR detectors can not be replaced by threshold detectors.

\appendix

\section{Photon-subtraction \label{phosub}}
The standard photon-subtraction probability is obtained from a high transmission beamsplitter where $U(\theta) = \exp [\theta (\hat{a}^{\dag} \hat{b} - \hat{a} \hat{b}^{\dag})]$ with $\theta <<1$. In this limit we can expand the unitary operator to the first-order in gate strength to obtain $1 + \theta (\hat{a}^{\dag} \hat{b} - \hat{a} \hat{b}^{\dag})$. For an arbitrary state and vacuum state incident on the beamsplitter, we obtain $U(\theta) |\psi\rangle \ket{0}  = N \left(|\psi\rangle \ket{0} - \theta \hat{a} \ket{\psi} \ket{1}\right) $, where $N= \sqrt{1 + \theta^2 \langle \hat{a}^{\dag} a  \rangle_{\psi}} $. If we measure a single photon in the second mode, the probability of a photon-subtraction on the input state is then given by $\theta^2 N^{-2}$. For beamsplitters with around $98\%$ transmission strength~\cite{Dakna1999,Gerrits2010}, this results in a probability $\sim 10^{-2}$. The resource states such as the one we consider require three successive photon-subtractions which results in a net probability of $10^{-6}$. If we use this latter probability in Eq.~\eqref{nmin} with $\epsilon \sim 10^{-3}$, we find that the required number of quantum gadgets is of the order of $10^5$ which is unfeasible.

\section{Numerical techniques \label{supp:num-tech}}
We discuss two optimization functions that were used for training the circuit parameters.  The landscape of our loss functions usually have several local minima, which makes it hard for standard local optimization methods because there is a very strong dependency on the initial conditions. The first global search function that we used is called \textsf{basinhopping} which is described in Algorithm \ref{alg: BH}.
\textsf{Basinhopping} is a stochastic algorithm which attempts to find the global minimum of a smooth scalar function~\cite{wales1997global}. The implementation we used in our simulation is from the \texttt{scipy} package.

\begin{algorithm}[H]
	\caption{\textsf{basinhopping}}
	\begin{algorithmic}[0]
		\Procedure{basinhopping}{$\mb{x}_0$, {\em f}, {\em args},  {\em niter}, {\em step\_size=1}}:
		\State $\mb{x}_\text{old}$ $\gets$ \texttt{local}\_\texttt{search}(\texttt{loss}, $\mb{x}_0$)
		\For{i in \texttt{range}({\em iter})}
		\State $\Delta$ $\gets $ \texttt{random}$(0,1)$
		\State $\mb{x}_{\text{jump}}\gets\mb{x}_{\text{old}} +$ {\em {step\_size}} $\times\Delta$
		\State $\mb{x}_{\text{new}}$ $\gets$ \texttt{local}\_\texttt{minimize}( $\mb{x}_{\text{jump}}$, \texttt{loss}, {\em args})
		\If{\texttt{accept}($\mb{x}_{\text{new}}$, $\mb{x}_{\text{old}}$)== True}
		\State $\mb{x}_{\text{old}} \gets \mb{x}_{\text{new}}$
		\EndIf
		\EndFor
		\EndProcedure
	\end{algorithmic}
	\label{alg: BH}
\end{algorithm}

The algorithm is iterative with each cycle composed of the following features:
\begin{enumerate}
    \item initialize the variables $\mb{x}_0$
    \item perform \textsf{local\_minimize}  to minimize $f(\mb{x}, args)$ starting from $\mb{x}_0$, to reach a local minimum we call $\mb{x}_\text{old}$
    \item randomly change the position of $\mb{x}_\text{old}$ with a tunable {\em step\_size}
    \item perform \textsf{local\_minimize} starting from $\mb{x}_\text{old}$ to reach a local minimum we call $\mb{x}_\text{new}$
    \item perform an acceptance test \texttt{accept}($\mb{x}_{\text{new}}$,$\mb{x}_{\text{old}}$): a simple rule could be that if $f(\mb{x}_\text{new}, args)<f(\mb{x}_\text{old}, args)$, $\mb{x}_\text{old} = \mb{x}_\text{new}$. However, the commonly used acceptance test is stochastic so as to maximize the likelihood of finding the global minimum. In the \texttt{scipy} implementation, the acceptance test used there is the Metropolis criterion of standard Monte Carlo algorithms~\cite{li1987monte}, where the probability of acceptance is given by $\exp[-(f(\mb{x}_\text{old}, args)-f(\mb{x}_\text{new}, args))/T]$. Here $T$ is a fictitious temperate to control the degree of randomness
    \item go back to step 2 and repeat this process  {\em niter} times
\end{enumerate}

This global minimization method has been shown to be extremely efficient for a wide variety of problems in physics and chemistry. For a stochastic global heuristic there is no way to determine if the true global minimum has actually been found. In our simulation we set {\em niter} $= 40$ , which is tuned to be able to have reproducible results. {\em step\_size} is set to be the default value from the \texttt{scipy} package. The algorithm for \textsf{local\_minimize} can in principle  be one of many options. In our simulation we choose sequential least squares programming (\textsf{SLSP})~\cite{bonnans2006numerical} which seems to be the fastest one for our task among the local search algorithms available in \texttt{scipy}. When there are no constraints on the variables, \textsf{SLSP} reduces to the well-known Newton's method~\cite{bonnans2006numerical}.\\

\noindent {\bf Remark.} In principle the $\mathsf{local\_search}$ function mentioned in Algorithm\,\ref{alg: main} can be any local minimization algorithm and in our simulation we use the well-known \textsf{BFGS} algorithm which is available from the \texttt{scipy} package. In principle, the \textsf{local\_search} and the \textsf{local\_minimize} algorithms can be the same. But we use different names due to the role they play in our  Algorithm.~\ref{alg: main}.

\subsection{Second-optimization over probability}
For our state preparation task, solely optimizing  fidelity to the target state is insufficient. To have a better architecture the conditional probability of preparing the trained state needs to be sufficiently high as well. One possible strategy to achieve both high fidelity and probability is to train the circuit to maximize fidelity and then use that point as a seed to further optimize the probability. This is exactly what we did for the three-mode case (see Algorithm \ref{alg: main}), where we found that the second optimization over probability did little harm to the fidelity.  

However, this is not the case for two-mode circuit where we found that second optimization quickly deteriorates the pre-trained high fidelity. To tackle this problem, making considerations for the fact that the computation overhead for training two-mode circuit is moderate, we did a brute-force optimization over probability. That is, we repeat the \textsf{basinhopping} {\em{nbh}} times. We then pick out the global optimum, trained to optimize fidelity with the highest probability. In our simulation, we found that using {\em nbh} = 20 and {\em niter} = 30 is enough to obtain reproducible results. This procedure is summarized in Algorithm \ref{al:opt_prob}.

\begin{algorithm}[H]
	\caption{optimization of probability for the two-mode architecture\label{al:opt_prob}}
	\begin{algorithmic}[0]
		\Procedure{prob\_opt}{{\em nbh}}:
		\State \textbf{initialize} $\mb{x}$\_list, prob\_list $\gets$ empty list
		\For{e in \texttt{range}({\em nbh})}
		\State \textbf{initialize} $\mb{x}_0$
		\State $\mb{x}\gets$ \texttt{basinhopping}($\mb{x}_0$, \texttt{loss}, {\em args}, {\em niter})
		\State \_ , prob, \_ , \_ $\gets$ \texttt{objective}($\mb{x}$)
		\State $\mb{x}$\_list.append($\mb{x}$)
		\State prob\_list.append(prob)
		\EndFor
		\State MaxIndex $\gets$ \texttt{max}(prob\_ls)
		\State \textbf{save} $\mb{x}$\_list[MaxIndex]
		\EndProcedure
	\end{algorithmic}
\end{algorithm}

\begin{acknowledgements}
We thank  Nathan Killoran,  Koushik Ramachandran, Maria Schuld, Kamil Bradler, Casey Myers, Juan Miguel Arrazola, and Brajesh Gupt for useful discussions. All numerical simulations were performed with open source software: the \textsf{StrawberryFields} quantum simulator\,\cite{killoran2018strawberry} is available at  \href{https://github.com/XanaduAI/strawberryfields}{https://github.com/XanaduAI/strawberryfields}; our Python code for  optimization built on \textsf{StrawberryFields} and \textsf{scipy} package~\cite{scipy}  can be accessed at \href{https://github.com/XanaduAI/constrained-quantum-learning}{https://github.com/XanaduAI/constrained-quantum-learning}.
\end{acknowledgements}

\bibliography{cubicML}

\end{document}

\begin{figure}
    \centering
    \includegraphics[scale=0.45]{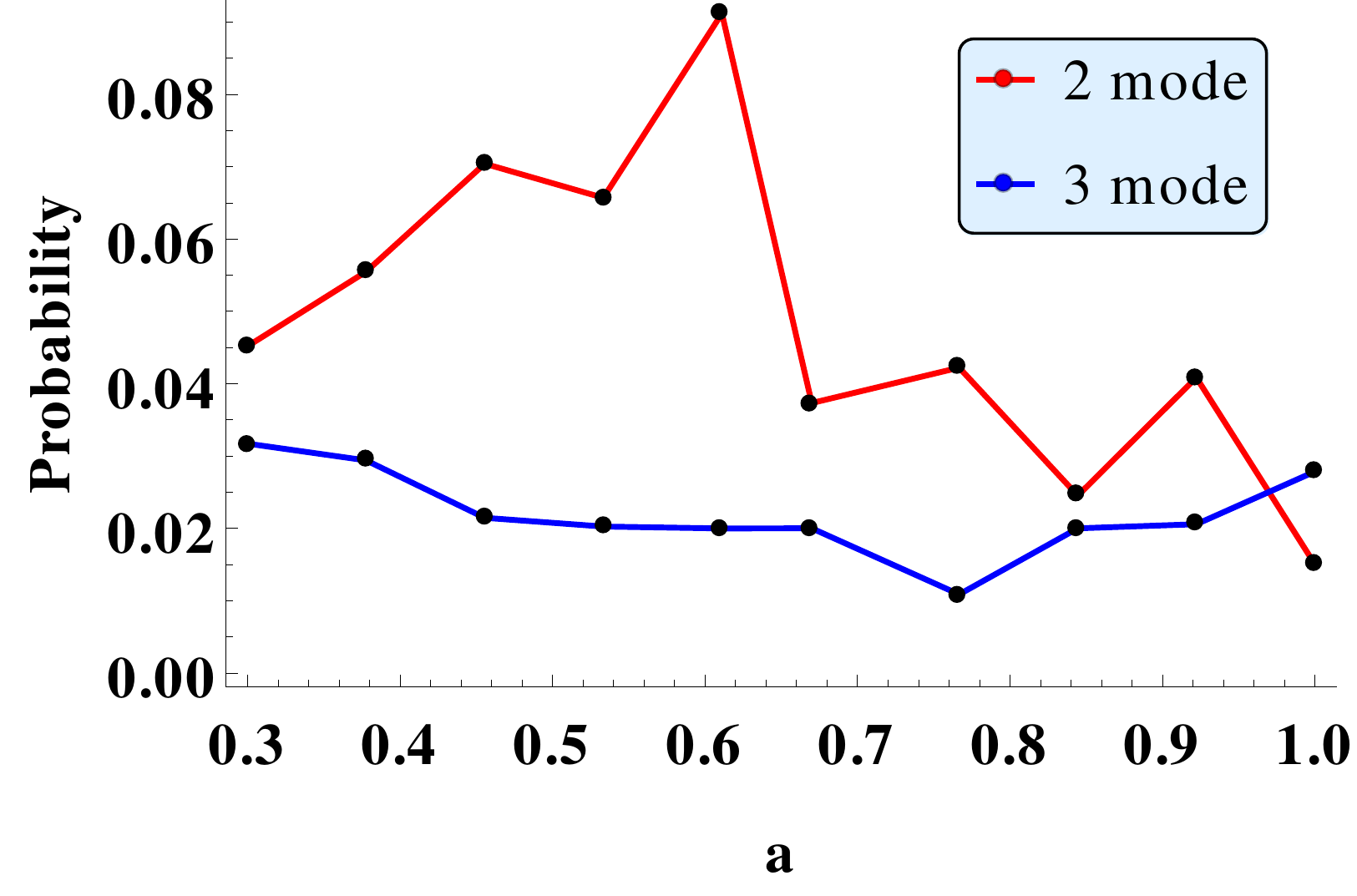}
    \caption{Probability of preparing the trained states using the two-mode and three-mode architectures for various values of target state parameter $a$. Data here is obtained from Algorithm.~1 (of the main paper) with {\em niter} = 40 and Algorithm.~\ref{al:opt_prob} with ({\em niter} = 30, {\em nbh} = 20). We find that for most cases the two-mode case performs better than the three-mode case, however, with a cost in the fidelity as mentioned in Fig.~3 of the main paper.}
    \label{fig:prob}
\end{figure}